\documentclass{article}
\usepackage{amsmath}
\usepackage{ws-rv-van}     

\newcommand{\address}[1]{\date{\emph{#1}}\maketitle}
\newcommand{\chapter}[2][]{\title{#2}}
\newcommand{\body}{}

\begin{document}

\chapter[A Projection to the Pure Spinor Space]{A Projection to the \\ Pure Spinor Space}\label{ra_ch1}

\author{Sebastian Guttenberg
}

\address{CAMGSD, Departamento de Matem\'atica,\\ Instituto Superior T\'ecnico,\\Av. Rovisco Pais 1,\\1049-–001 Lisboa, Portugal.\\ sgutten@math.ist.utl.pt
}

\begin{abstract}
This article is based on a talk given at the \emph{Memorial Conference for
Maximilian Kreuzer} at the ESI in Vienna and contains a compact summary of a recent collaboration with P.A. Grassi. A non-linear projection from the space of SO(10) Weyl spinors to the space of pure spinors is presented together with some of its particular properties. This projection can be used to remove the constraints from Berkovits' pure spinor superstring while introducing additional gauge symmetries. This should allow to make transitions to equivalent formulations which might shed light on the origin of the pure spinor ghosts. It might also be useful in the context of path integral measures for the pure spinor string. 
\end{abstract}

\body

\section{Introduction}\label{ra_sec1}
As this article is a contribution to a memorial volume for Maximilian Kreuzer, let me first put it into context. From 2003 to 2007, Max Kreuzer  was the supervisor for my PhD thesis which mainly addressed the pure spinor formulation of string theory and aspects of generalized complex geometry. I am immensely grateful that he convinced me of the importance of these fields and for his  most valuable initial input.
 Although both subjects were not part of his main research areas at that time, they were always important to him, and I will sketch below shortly, why. After this small detour, this article will concentrate on a small aspect within just one of the fields, namely within the pure spinor string.

In particular among string theorists Max Kreuzer was mainly known for his major contributions to classifications of Calabi-Yau spaces. These spaces have been extremely popular in string theory over the last decades, as compactifications on them were known to lead to effective supersymmetry in four dimensional spacetime. However, already from the early times \cite{Strominger:1986uh} it was well known that more general cases can also lead to supersymmetry in four dimensions. Nevertheless for about two decades research was focusing on Calabi-Yaus, because they were mathematically treatable and revealed fascinating properties such as mirror-symmetry. Even when flux-compactification became more fashionable, the fluxes were studied in the beginning mainly without backreactions, thus not destroying the Ricci-flatness. The focusing on Calabi-Yaus thus went so far that in particular many young researchers were not even aware anymore of the possibility to have supersymmetry without Calabi-Yaus. Max Kreuzer, in spite of having specialized on Calabi-Yau spaces, was not only aware of this fact, but was following with great interest every development which would make the non-Calabi-Yau cases accessible. 

One important development was Berkovits' invention of a covariantly quantized and still manifestly target space supersymmetric formulation of the superstring\cite{Berkovits:2000fe}. The new formalism was important in this context, because non-Calabi-Yau compactifications with effective fourdimensional supersymmetry need to contain nonvanishing fluxes, in particular they might contain so-called RR-fluxes. It is still not known how to couple these RR-fluxes to the standard RNS-formulation of string theory. This means that any complete study of RR-flux-compactifications with back-reactions from the worldsheet point of view has to use a worldsheet description which allows the coupling to RR-fields. The first obvious candidate, the Green Schwarz formalism, has this property, but it was in turn not possible to covariantly quantize it. The pure spinor formulation (which can only to some extent be seen as a covariant quantization of the Green Schwarz string) has neither of the problems and is therefore the first serious candidate for a full quantum study of flux-backgrounds. 

The next important development was the observation\cite{Grana:2004bg,Grana:2004??,Jeschek:2004wy} that four-\linebreak[4]dimensional spacetime supersymmetry forces the compactification manifold to be a generalized Calabi-Yau manifold in the sense of Hitchin\cite{Hitchin:2004ut}. This gives a nice geometric generalization of the previous relation between supersymmetry and ordinary Calabi-Yau manifolds in the absence of fluxes. The tools of generalized complex geometry had become quite powerful and for the first time it seemed that a systematic study of general supersymmetric flux compactifications would become a reachable long-term goal. Max Kreuzer realized the importance of these developments and convinced me to work on them. In that sense he is also at the origin of the work that will be presented here. 

Interestingly both topics, the pure spinor string as well as generalized complex geometry make heavy use of the notion of a "pure spinor".
A pure spinor can be seen as a possible vacuum for a Fock-space representation of spinors. Or in other words, it is defined to be annihilated by a subspace  which has half the dimension of the Clifford-vector space (spanned by the Dirac-Gamma-matrices). At present, the fact that pure spinors appear in both discussed subjects seems more like a coincidence, because in the pure spinor string it is an SO(10) (or SO(1,9)) pure spinor, while in generalized complex geometry it is (for a 6-dimensional compactification manifold) an SO(6,6) pure spinor. Although it would be interesting to study at least the interplay of these two pure spinors when the pure spinor string is coupled to a generalized complex background, this article will concentrate on the SO(10) pure spinor $\lambda^\alpha$ of Berkovits' formulation of string theory. In this latter case, the pure spinor definition can be rewritten in a Lorentz covariant way as a set of quadratic constraints on the spinor:
\begin{equation}(\lambda\gamma^a\lambda)=0\end{equation}

For some fundamental calculations, the pure spinor constraint had been explicitly solved in a U(5) covariant parametrization\cite{Berkovits:2000fe} which is based on the Fock-space representation of spinors. As mentioned above, this Fock space representation uses a vacuum which is itself a pure spinor. This means, when solving the pure spinor constraint in this parametrization, one expresses a general pure spinor in terms of a particular one (the vacuum). 
It is thus a natural conceptional question how to obtain a pure spinor to start with. Or alternatively how to construct a pure spinor without a vacuum. This question was answered in an article together with P.A. Grassi \cite{Grassi:2011ie} by presenting a family of projections to the pure spinor space. The present contribution will be a summary of that rather technical article by focusing on one interesting representative of that family. 

Apart from the purely conceptional interest, there might also be important applications of this projection. In spite of remarkable progress in pure spinor string theory in recent years, the appearance of a pure spinor is still a bit mysterious. Several approaches have related the pure spinor formalism to other formalisms in various ways \cite{SorokinMatone:2002ft,Berkovits:2004tw,Gaona:2005yw,Berkovits:2007wz,Aisaka:2005vn,Oda:2011kh,Hoogeveen:2007tu}, certainly providing important insight. However, obtaining the pure spinor ghost directly from the gauge fixing of a fermionic gauge symmetry seems not possible, because the constraint on the ghost is quadratic and cannot be directly translated into an equivalent constraint for the fermionic gauge parameter.
By replacing in the pure spinor action the pure spinor by the projection of a general spinor, one can remove the constraint while introducing additional gauge symmetries. One can try to find different fixings of these gauge symmetries which might allow the interpretation of the ghosts as coming from a classical gauge symmetry. In addition the projection might be valuable for constructing path-integral measures. 

\section{A non-linear projection to the pure spinor space}

One can define a family of projections\cite{Grassi:2011ie} $P_{(f)}^\alpha$ parametrized by real-valued functions $f$ and mapping from the space of SO(10) Weyl spinors onto the space of pure SO(10) Weyl spinors: 
\begin{eqnarray}
P_{(f)}^{\alpha}(\rho,\bar{\rho})&\equiv & f\!\bigl(\!\tfrac{(\rho\gamma^{a}\rho)(\bar\rho\gamma_{a}\bar\rho)}{2(\rho\bar{\rho})^2} \!\bigr)\Bigl(\rho^{\alpha}-\tfrac{1}{2}\tfrac{(\rho\gamma^{a}\rho)(\bar{\rho}\gamma_{a})^{\alpha}}{(\rho\bar{\rho})+\sqrt{(\rho\bar{\rho})^{2}-\frac{1}{2}(\rho\gamma^{b}\rho)(\bar{\rho}\gamma_{b}\bar{\rho})}}\Bigr)\label{Pf}\\
&&{\rm with\;}f(0)\equiv 1 \nonumber
\end{eqnarray}
It is obvious that for every $f$ with $f(0)=1$ the map $P_{(f)}^\alpha$ reduces to the identity-map if $\rho^\alpha$ is a pure spinor, i.e. if $\rho\gamma^c\rho$ vanishes. Most of the proofs of facts presented in this summarizing article will be omitted and can be found in the original article\cite{Grassi:2011ie}. However, at least the calculation for the main statement, namely that the image $P_{(f)}^\alpha(\rho,\bar\rho)$ is a pure spinor for every Weyl spinor $\rho^\alpha$ will be sketched in the following lines. To this end, let us consider the corresponding bilinear (neglecting the overall factor $f^2$):
\begin{eqnarray}
\lefteqn{P_{(f)}^{\alpha}(\rho,\bar{\rho})\gamma_{\alpha\beta}^{c}P_{(f)}^{\beta}(\rho,\bar{\rho})\propto}\nonumber&&\\
&\propto&(\rho\gamma^{c}\rho)-\tfrac{1}{(\rho\bar{\rho})+\sqrt{(\rho\bar{\rho})^{2}-\frac{1}{2}(\rho\gamma^{b}\rho)(\bar{\rho}\gamma_{b}\bar{\rho})}}\underbrace{(\rho\gamma^{a}\rho)(\bar{\rho}\gamma_{a}\gamma^{c}\rho)}_{{\scriptscriptstyle 1}\!\!\!\!\bigcirc}+\nonumber\\
&&+\tfrac{1}{4}\tfrac{(\rho\gamma^{a}\rho)}{(\rho\bar{\rho})+\sqrt{(\rho\bar{\rho})^{2}-\frac{1}{2}(\rho\gamma^{d}\rho)(\bar{\rho}\gamma_{d}\bar{\rho})}}\underbrace{(\bar{\rho}\gamma_{a}\gamma^{c}\gamma_{b}\bar{\rho})}_{{\scriptscriptstyle 2}\!\!\!\!\bigcirc}\tfrac{(\rho\gamma^{b}\rho)}{(\rho\bar{\rho})+\sqrt{(\rho\bar{\rho})^{2}-\frac{1}{2}(\rho\gamma^{e}\rho)(\bar{\rho}\gamma_{e}\bar{\rho})}}\qquad 
\end{eqnarray}
For the term $\; {\scriptstyle{\scriptscriptstyle 1}\!\!\!\!\bigcirc}$, in order to make use of the Fierz identity 
\mbox{$(\rho\gamma^{a}\rho)\!(\gamma_{a}\rho)_\alpha\!=0$}, we need to reorder the gamma-matrices via the Clifford algebra \mbox{$\gamma_a\gamma^c=-\gamma^c\gamma_a+2\delta_a^c$}.
Also for the term $\; {\scriptstyle{\scriptscriptstyle 2}\!\!\!\!\bigcirc}$ it helps to reorder the Gamma-matrices $\gamma_a$ and $\gamma^c$ via this relation, because the resulting product $\gamma_a\gamma_b$ gets symmetrized by the contractions and thus reduces to $\eta_{ab}$. In this way, one obtains
\begin{eqnarray}
\lefteqn{P_{(f)}^{\alpha}(\rho,\bar{\rho})\gamma_{\alpha\beta}^{c}P_{(f)}^{\beta}(\rho,\bar{\rho})\propto}\nonumber&&\\
&\propto&(\rho\gamma^{c}\rho)-\tfrac{1}{(\rho\bar{\rho})+\sqrt{(\rho\bar{\rho})^{2}-\frac{1}{2}(\rho\gamma^{b}\rho)(\bar{\rho}\gamma_{b}\bar{\rho})}}
2(\rho\gamma^{c}\rho)(\bar{\rho}\rho)+\nonumber\\
&&+\tfrac{1}{2}\tfrac{(\rho\gamma^{c}\rho)}{(\rho\bar{\rho})+\sqrt{(\rho\bar{\rho})^{2}-\frac{1}{2}(\rho\gamma^{d}\rho)(\bar{\rho}\gamma_{d}\bar{\rho})}}
(\bar{\rho}\gamma_{b}\bar{\rho})\tfrac{(\rho\gamma^{b}\rho)}{(\rho\bar{\rho})+\sqrt{(\rho\bar{\rho})^{2}-\frac{1}{2}(\rho\gamma^{e}\rho)(\bar{\rho}\gamma_{e}\bar{\rho})}}+\qquad \nonumber\\
&&-\tfrac{1}{4}\tfrac{(\rho\gamma^{a}\rho)}{(\rho\bar{\rho})+\sqrt{(\rho\bar{\rho})^{2}-\frac{1}{2}(\rho\gamma^{d}\rho)(\bar{\rho}\gamma_{d}\bar{\rho})}}
(\bar{\rho}\gamma^{c}\bar{\rho})\tfrac{(\rho\gamma_a\rho)}{(\rho\bar{\rho})+\sqrt{(\rho\bar{\rho})^{2}-\frac{1}{2}(\rho\gamma^{e}\rho)(\bar{\rho}\gamma_{e}\bar{\rho})}}\qquad 
\end{eqnarray}
The last summand vanishes due to the Fierz identity \mbox{$\!(\rho\gamma^{a}\rho)\!(\gamma_{a}\rho)_\alpha\!\!=\!0$}. Putting the second and third summand on a common denominator\linebreak[4] \mbox{$\scriptstyle\bigl((\rho\bar{\rho})+\sqrt{(\rho\bar{\rho})^{2}-\frac{1}{2}(\rho\gamma^{b}\rho)(\bar{\rho}\gamma_{b}\bar{\rho})}\bigr)^2=2(\rho\bar{\rho})^{2}-\frac{1}{2}(\rho\gamma^{b}\rho)(\bar{\rho}\gamma_{b}\bar{\rho})+2(\rho\bar{\rho})\sqrt{(\rho\bar{\rho})^{2}-\frac{1}{2}(\rho\gamma^{b}\rho)(\bar{\rho}\gamma_{b}\bar{\rho})}$}, their\linebreak[4] sum simplifies to $-(\rho\gamma^{c}\rho)$ and thus precisely cancels the first summand, such that indeed $P_{(f)}^{\alpha}(\rho,\bar{\rho})$ is a pure spinor: 
\begin{equation}P_{(f)}^{\alpha}(\rho,\bar{\rho})\gamma_{\alpha\beta}^{c}P_{(f)}^{\beta}(\rho,\bar{\rho})=0\end{equation}
The denominator with the square root in the projection (\ref{Pf}) is suprisingly well defined, because it can be shown\cite{Grassi:2011ie} that 
for all Weyl spinors $\rho^\alpha$ 
\begin{equation}(\rho\bar{\rho})^2\geq\tfrac 12(\rho\gamma^{b}\rho)(\bar\rho\gamma_{b}\bar\rho) \end{equation}
Let us define the Jacobian matrix 
\begin{equation}\left(\begin{array}{cc}
\Pi_{(f)\bot\beta}^{\alpha}(\rho,\bar{\rho}) & \pi_{(f)\bot}^{\alpha\beta}(\rho,\bar{\rho})\\
\bar{\pi}_{(f)\bot\alpha\beta}(\rho,\bar{\rho}) & \bar{\Pi}_{(f)\bot\alpha}{}^{\beta}(\rho,\bar{\rho})
\end{array}\right)\equiv\left(\begin{array}{cc}
\partial_{\rho^{\beta}}P_{(f)}^{\alpha}(\rho,\bar{\rho}) & \partial_{\bar{\rho}_{\beta}}P_{(f)}^{\alpha}(\rho,\bar{\rho})\\
\partial_{\rho^{\beta}}\bar{P}_{(f)\alpha}(\rho,\bar{\rho}) & \partial_{\bar{\rho}_{\beta}}\bar{P}_{(f)\alpha}(\rho,\bar{\rho})
\end{array}\right)\end{equation}
The subscripts $\scriptstyle\bot$ are a reminder that the Jacobian matrix linearly maps spinors to a subspace which is '$\gamma$-orthogonal' to the pure spinor \mbox{$\lambda^\alpha\equiv P^\alpha_{(f)}(\rho,\bar\rho)$}. In particular the variation of the pure spinor $\lambda^\alpha$ which is given by\linebreak[4] $\delta\lambda^\alpha=(\Pi_{(f)\bot}(\rho,\bar\rho)\delta\rho)^\alpha+(\pi_{(f)\bot}(\rho,\bar\rho)\delta\bar\rho)^\alpha$ is $\gamma$-orthogonal to $\lambda^\alpha$ in the sense \mbox{$(\lambda\gamma^a\delta\lambda)=0$}. For the Jacobian matrix itself this means
\begin{eqnarray}
P_{(f)}^{\alpha}(\rho,\bar\rho)\gamma_{\alpha\gamma}^{c}\Pi_{(f)\bot\beta}^{\gamma}(\rho,\bar{\rho})&=&0\\P_{(f)}^{\alpha}(\rho,\bar\rho)\gamma_{\alpha\gamma}^{c}\pi_{(f)\bot}^{\gamma\beta}(\rho,\bar{\rho})&=&0\qquad\forall\rho
\end{eqnarray}
Independent of the choice for $f$, the Jacobian matrix reduces on the constraint surface $\rho^\alpha=\lambda^\alpha$, where $\lambda^\alpha$ is pure, to 
\begin{equation}\left(
\begin{array}{cc}
\Pi_{(f)\bot}(\lambda,\bar{\lambda}) & \pi_{(f)\bot}(\lambda,\bar{\lambda})\\
\bar{\pi}_{(f)\bot}(\lambda,\bar{\lambda}) & \bar{\Pi}_{(f)\bot}(\lambda,\bar{\lambda})
\end{array}
\right)=\left(
\begin{array}{cc}
\Pi_{\bot} & 0\\
0 & \bar{\Pi}_{\bot}
\end{array}
\right)\end{equation}
where $\Pi_{\bot}$ is a Hermitean linear projector given by
\begin{equation}
\Pi_{\bot\beta}^{\alpha}=\delta_{\beta}^{\alpha}-\tfrac{1}{2}\tfrac{(\gamma_{a}\bar{\lambda})^{\alpha}(\lambda\gamma^{a})_{\beta}}{\overset{}{(\lambda\bar{\lambda})}}
\end{equation}
The transpose of this projector appears in the pure spinor string literature\cite{Oda:2001zm,Berkovits:2010zz} in the context of the gauge invariant part of the antighost $\omega_{z\alpha}$ and is  there denoted as $1-K$: 
\begin{equation}\tilde\omega_{z\alpha}=(\Pi_{\bot}^{T}\omega_{z})_{\alpha}\quad\rm{(gauge\, invariant)}\end{equation}
There is one choice $f=h$ with 
\begin{equation}h(\xi)\equiv\tfrac{1+\sqrt{1-\xi}}{2\sqrt{1-\xi}}\end{equation}
for which the Jacobian is also Hermitean off the constraint surface
\begin{equation}
\left(\begin{array}{cc}
\Pi^\dag_{(h)\bot}(\rho,\bar{\rho}) & \pi^T_{(h)\bot}(\rho,\bar{\rho})\\
\bar{\pi}^T_{(h)\bot}(\rho,\bar{\rho}) & \Pi^T_{(h)\bot}(\rho,\bar{\rho})
\end{array}\right)=
\left(\begin{array}{cc}
\Pi_{(h)\bot}(\rho,\bar{\rho}) & \pi_{(h)\bot}(\rho,\bar{\rho})\\
\bar{\pi}_{(h)\bot}(\rho,\bar{\rho}) & \bar{\Pi}_{(h)\bot}(\rho,\bar{\rho})
\end{array}\right)
\end{equation}
For this choice $f=h$, the non-linear projection (\ref{Pf}) takes the form
\begin{eqnarray}
P_{(h)}^{\alpha}(\rho,\bar{\rho})&\equiv &\tfrac{(\rho\bar{\rho})+\sqrt{(\rho\bar{\rho})^{2}-\frac{1}{2}(\rho\gamma^{a}\rho)(\bar{\rho}\gamma_{a}\bar{\rho})}}{2\sqrt{(\rho\bar{\rho})^{2}-\frac{1}{2}(\rho\gamma^{b}\rho)(\bar{\rho}\gamma_{b}\bar{\rho})}}\rho^{\alpha}-\tfrac{(\rho\gamma^{a}\rho)}{4\sqrt{(\rho\bar{\rho})^{2}-\frac{1}{2}(\rho\gamma^{b}\rho)(\bar{\rho}\gamma_{b}\bar{\rho})}}(\bar{\rho}\gamma_{a})^{\alpha}\qquad 
\end{eqnarray}
It turns out that this case has some additional nice properties. 
In particular, there exists a potential 
\begin{equation}
\Phi(\rho,\bar{\rho})\equiv\tfrac{(\rho\bar{\rho})}{2}\Bigl(1+\sqrt{1-\tfrac{(\rho\gamma^{a}\rho)(\bar\rho\gamma_{a}\bar\rho)}{2(\rho\bar{\rho})^2}}\Bigr)
\end{equation}
 such that 
\begin{equation}
P_{(h)}^{\alpha}=\partial_{\bar{\rho}_{\alpha}}\Phi\quad,\quad\bar{P}_{(h)\alpha}=\partial_{\rho^{\alpha}}\Phi 
\end{equation}
Furthermore this potential $\Phi$  can be written as
\begin{equation}\Phi(\rho,\bar{\rho})=P_{(h)}^{\alpha}(\rho,\bar{\rho})\bar{P}_{(h)\alpha}(\rho,\bar{\rho}) \end{equation}
It is therefore related to the pullback of the pure spinor K\"ahler potential along the projection $P_{(h)}^{\alpha}$ into the ambient space. 

In order to observe one more interesting property of $P_{(h)}^{\alpha}$, let us regard it as part of a variable transformation, by introducing additional variables (those which are projected out). For consistency with the original article\cite{Grassi:2011ie} these variables will be called $\check\zeta^a$. So altogether we consider the following variable transformation:
\begin{eqnarray}
(\rho^{\alpha},\bar{\rho}_{\alpha})&\mapsto&(\lambda^{\alpha},\bar{\lambda}_{\alpha},\check\zeta^{a},\bar{\check\zeta}_{a})\\
\mbox{with }\lambda^{\alpha}&\equiv&P_{(h)}^{\alpha}(\rho,\bar{\rho})\\
\check\zeta^{a}&\equiv&\tfrac{1}{2}(\rho\gamma^{a}\rho)
\end{eqnarray}
It is obvious that $\check\zeta^{a}$ captures the non-pure-spinor part of $\rho^\alpha$. The statement is thus that every Weyl spinor $\rho^\alpha$ can be (redundantly) parametrized by a pure spinor $\lambda^\alpha$ and the variable $\check\zeta^{a}$. So far this is not surprising. What is nice about the case $f=h$, is that the inverse variable transformation takes a very simple form: 
\begin{equation}\rho^{\alpha}=\lambda^{\alpha}+\tfrac{1}{2}\frac{\check\zeta^{a}(\bar{\lambda}\gamma_{a})^{\alpha}}{(\lambda\bar{\lambda})}\end{equation}
Note that like the pure spinor $\lambda^\alpha$ also the variable $\check\zeta^{a}$ is not free, but obeys some constraints which remove 5 of its 10 degrees of freedom:
\begin{equation}\check\zeta^{a}(\lambda\gamma_{a})_{\alpha}=\check\zeta^{a}\check\zeta_{a}=0\end{equation}
One possible application of this variable transformation is to start from a volume form in the ambient space (Weyl spinors) and see if this volume form  decomposes after the variable transformation into a $\lambda^\alpha$ volume form and a $\check\zeta^a$ volume form. This might be a way to rederive the holomorphic tree-level volume form for the pure spinor partition function and perhaps also to derive loop path integral measures. However, already at tree level, complete factorization of the promising result\cite{Grassi:2011ie}
\begin{eqnarray}
[d^{16}\rho]|_{\check{\zeta}^{a}=0}\!&\propto&\!\tfrac{1}{(\overset{}{\lambda}\bar{\lambda})^{3}}\gamma_{bcd}^{\alpha_{1}\alpha_{2}}(\bar{\lambda}\gamma_{a_{1}})^{\alpha_{3}}\cdots(\bar{\lambda}\gamma_{a_{3}})^{\alpha_{5}}\epsilon_{\alpha_{1}\ldots\alpha_{16}}\boldsymbol{\rm{d}}\!\lambda^{\alpha_{6}}\cdots\boldsymbol{\rm{d}}\!\lambda^{\alpha_{16}}\!\!\times\nonumber\\
&&\!\times\!\tfrac{1}{(\overset{}{\lambda}\bar{\lambda})^{2}}(\bar{\lambda}\gamma^{bcd}{}_{a_{4}a_{5}}\bar{\lambda})\boldsymbol{\rm{d}}\!\check{\zeta}^{a_{1}}\cdots\boldsymbol{\rm{d}}\!\check{\zeta}^{a_{5}}
\end{eqnarray}
 would require some non-trivial identity which has not yet been verified. 

Another application of the projection $P_{(h)}^\alpha$ is to remove the pure spinor constraint from the pure spinor string while introducing additional gauge symmetries. The idea is very simple: Having an action $S[\lambda,\bar\lambda]$ depending on the pure spinor $\lambda^\alpha$, it suffices to replace this pure spinor by $P_{(h)}^\alpha(\rho,\bar\rho)$ and consider the action to be a functional of the unconstrained variable $\rho^\alpha$:
\begin{equation}
\tilde S[\rho,\bar\rho]\equiv S[P_{(h)}(\rho,\bar\rho),\bar P_{(h)}(\rho,\bar\rho)]
\end{equation}
Obviously any variation of $\rho^\alpha$ which will not change the projection $P_{(h)}^\alpha(\rho,\bar\rho)$ will be an additional symmetry of the new action. Remember the defining relation for the Jacobian matrix:
\begin{equation}
\left(\!\!\begin{array}{c}
\delta P_{(h)}^{\alpha}(\rho,\bar{\rho})\\
\delta\bar{P}_{(h)\alpha}(\rho,\bar{\rho})
\end{array}\!\!\right)\!\equiv\!\left(\!\!\begin{array}{cc}
\Pi_{(h)\bot\beta}^{\alpha}(\rho,\bar{\rho}) & \pi_{(h)\bot}^{\alpha\beta}(\rho,\bar{\rho})\\
\bar{\pi}_{(h)\bot\alpha\beta}(\rho,\bar{\rho}) & \bar{\Pi}_{(h)\bot\alpha}{}^{\beta}(\rho,\bar{\rho})
\end{array}\!\!\right)\!\!\left(\!\!
\begin{array}{c}
\delta\rho^{\beta}\\
\delta\bar{\rho}_{\beta}
\end{array}\!\!\right)\label{deltaP}
\end{equation}
On the constraint surface $\Pi_{\bot}$ is a proper projection matrix and thus obeys $\Pi_{\bot}^{2}=\Pi_{\bot}$  implying $\Pi_{\bot}\Pi_{\Vert}=0$ where  \begin{equation}
\Pi_{\Vert}\equiv 1-\Pi_{\bot}
\end{equation}
 This suggests that the additional symmetry transformations should be something of the form $\delta_{({\rm sym})}\rho\propto\Pi_{\Vert}$ . Off the constraint surface we do a priori not have $\Pi_{(h)\bot}^{2}(\rho,\bar\rho)=\Pi_{(h)\bot}(\rho,\bar\rho)$. However, one can derive a similar relation by taking the variation of the projection property $P_{(h)}(P_{(h)}(\rho,\bar\rho),\bar P_{(h)}(\rho,\bar\rho))=P_{(h)}(\rho,\bar\rho)$, namely $\Pi_{(h)\bot}(\lambda,\bar{\lambda})\Pi_{(h)\bot}(\rho,\bar{\rho})=\Pi_{(h)\bot}(\rho,\bar{\rho})$  with $\lambda\equiv P_{(h)}(\rho,\bar\rho))$. And this directly implies
\begin{equation}
\Pi_{(h)\Vert}(\lambda,\bar{\lambda})\Pi_{(h)\bot}(\rho,\bar{\rho})=0
\end{equation}
It is only because of the Hermiticity of $\Pi_{(h)\bot}(\rho,\bar{\rho})$ that the order in the above matrix-multiplication can be interchanged by simply taking the Hermitian conjugate of the whole equation: 
\begin{equation}
\Pi_{(h)\bot}(\rho,\bar{\rho})\underbrace{\Pi_{(h)\Vert}(\lambda,\bar{\lambda})}_{\Pi_{\Vert}}=0
\end{equation}
Comparing this with (\ref{deltaP}), it is now obvious that one obtains additional gauge symmetries of the form
\begin{equation}
\delta \rho^\alpha =(\Pi_{\Vert} \nu)^\alpha
\end{equation}
with some spinorial gauge parameter $\nu^\alpha$. 
In the full non-minimal formalism of the pure spinor superstring, there are additional constrained variables $\boldsymbol{r}_\alpha$, such that removing all constraints becomes slightly more involved, but following the same ideas. This analysis leads also to linear projectors to the gauge invariant parts of $\boldsymbol{s}_z^\alpha$ (the conjugate of $\boldsymbol{r}_\alpha$) and of $\bar w_{z\alpha}$ (the complex conjugate antighost):\cite{Grassi:2011ie} 
\begin{eqnarray}\tilde{\boldsymbol s}_{z}^{\alpha}&=&(\Pi_{\bot}\boldsymbol s_{z})^{\alpha}\quad\quad\rm{(gauge\, invariant)}\\
\tilde{\bar{\omega}}_{z}^{\alpha}&=&(\Pi_{\bot}\bar{\omega}_{z})^{\alpha}-(\Pi_{\bot}\gamma_{a}\boldsymbol r)^{\alpha}\frac{(\lambda\gamma^{a}\boldsymbol s_{z})}{2(\lambda\bar{\lambda})}\quad\rm{(gauge\, invariant)}
\end{eqnarray}
For this to be a linear projection, it has to be regarded as one single map acting on the tuple $(\boldsymbol s_{z}^\alpha,\bar\omega_{z}^{\alpha})$, instead of two separate maps. 

The above discussed additional gauge symmetry of course allows to fix $\rho^\alpha$ to be a pure spinor (thus returning to the original action), but in addition it allows to choose many different gauges. Two possible directions were sketched in the original article:\cite{Grassi:2011ie} 
\begin{eqnarray}
(\omega_{z}\gamma^{a}\bar{P}_{(h)}(\rho,\bar{\rho}))&\stackrel{!}{=}& 0\quad\mbox{(gauge 1)}
\end{eqnarray}
This can be regarded as a constraint on the antighost only, thus leaving the ghosts $\rho^\alpha$ unconstrained. However, the action becomes non-free in contrast to the starting point where the ghost $\lambda^\alpha$ was pure but the action was essentially free. One can force to obtain a free action also for a gauge fixing that involves the antighost. In its full glory it involves also the nonminimal variable $\boldsymbol{s}_z^\alpha$. But if we neglect this variable for simplicity, the corresponding gauge fixing reads
\begin{eqnarray}
\left(\!\begin{array}{cc}
\bar{\Pi}_{(h)\bot}(\rho,\bar{\rho}) & \bar{\pi}_{(h)\bot}(\rho,\bar{\rho})\\
\pi_{(h)\bot}(\rho,\bar{\rho}) & \Pi_{(h)\bot}(\rho,\bar{\rho})
\end{array}\!\!\right)\!\!\!\left(\!\!\begin{array}{c}
\omega_{z}\\
\bar{\omega}_{z}
\end{array}\!\!\right)&\!\stackrel{!}{=}\!&\!\left(\!\!\begin{array}{c}
\omega_{z}\\
\bar{\omega}_{z}
\end{array}\!\!\right)\quad\mbox{(gauge 2)}
\end{eqnarray}
It remains to study if there is any advantage of having a constraint on the antighost instead of having one on the ghost. Perhaps one can find also different interesting gauge fixings. Ideally they would be linear in all ghost variables in order to be able to transfer the constraint from the ghosts to the corresponding gauge parameters of opposite statistics. This would allow to derive these ghosts from an underlying gauge symmetry.   

\section{Conclusions}
The family of projections $P_{(f)}^\alpha$ from the space of Weyl spinors to the space of pure spinors (\ref{Pf}) allow to explicitly construct pure spinors in a covariant way and also to provide possible vacua for a Clifford representation. 
It has one representative $P_{(h)}^\alpha$ which has a Hermitian Jacobian matrix. This projection turned out to have a couple of interesting properties. Hermiticity itself was necessary to obtain the explicit form of the additional gauge symmetries which are obtained when the pure spinor in the action functional is replaced by the projection of an unconstrained Weyl spinor. This gauge symmetry allows to choose a different gauge which might lead to a formulation that sheds additional light on the origin of the pure spinor formalism. 
Seeing the projection as part of a variable transformation allows to study the transformation of volume forms and might give insight on the loop measure of the pure spinor partition function. 

\subsection*{Acknowledgements}
Once again many thanks to Maximilian Kreuzer for getting me interested into the pure spinor formulation of the superstring. Some of his ideas and points of view certainly had big impact on my work. Many thanks also to the organizers of the memorial conference for the kind invitation and -- well -- for organizing. In particular I would like to acknowledge some valuable comments of Anton Rebhan and Johanna Knapp on the first draft of this article.
This contribution was written with financial support of the Portuguese research foundation FCT (reference number SFRH/BPD/63719/2009).
\newpage 
\bibliographystyle{ws-rv-van}


\end{document}